\documentclass{svproc}
\usepackage{url}

\usepackage{color,xcolor}
\usepackage{latexsym,amsmath,amsfonts,amssymb,bbm,amsbsy}
\usepackage{epsfig,graphics,calc,graphicx,pict2e}
\usepackage{subcaption}

\newcommand*\diff{\mathop{}\!\mathrm{d}}

\begin{document}
\mainmatter              
\title{Pattern formation in three--state systems: Towards understanding morphology formation in the presence of evaporation}
%
\titlerunning{$\phantom.$}  
%
\author{Emilio N.M. Cirillo\inst{1} \and Rainey Lyons\inst{2}
\and Adrian Muntean\inst{2} \and Stela Andrea Muntean\inst{3}}
%
\authorrunning{$\phantom.$} 
%
%
\institute{Dipartimento di Scienze di Base e Applicate per l'Ingegneria (SBAI), Sapienza Universit\`a di Roma, Italy,\\
\email{emilio.cirillo@uniroma1.it}\\ 
\and
Department of Mathematics and Computer Science, Karlstad University, Sweden\\
\email{rainey.lyons@kau.se},
\email{adrian.muntean@kau.se}, 
\and
Department of Engineering and Physics, Karlstad University, Sweden\\
\email{andrea.muntean@kau.se}\\
}

\maketitle              

\begin{abstract}
Inspired by experimental evidence collected when processing thin films from ternary solutions made of two solutes, typically polymers, and one solvent, we computationally study the morphology formation of domains obtained in three-state systems using both a lattice model and a continuum counterpart. 
The lattice-based approach relies on the Blume--Capel nearest neighbor 
model with bulk conservative Kawasaki dynamics, whereas as continuum 
system we consider a coupled system of evolution equations that is derived as hydrodynamic limit when replacing the nearest neighbor interaction in the lattice case by a suitable Kac potential. 
We explore how the obtained morphology depends on the solvent content in the mixture. In particular, we study how these scenarios change when 
the solvent is allowed to evaporate.
\end{abstract}

\section{Introduction}
\label{intro}
We are interested in understanding how phase separation takes place in interacting ternary mixtures with one evaporating component. As our work is motivated by applications like building active layers for organic electronics, for instance, like organic solar cells (OSC) \cite{Moons_2007,Kouijzer_2013,Huang_2014,Schaefer_2016,Wodo,Ronsin_2022,Bruijn} and thin polymer composite layers for solution--borne adhesives \cite{Nils}, our focus lies on models describing the time and space evolution of polymer blends in solution where the solvent evaporates.  
In this paper, we use modeling and simulation tools to address the question: How are morphologies affected by variations in the solvent evaporation rate? Specifically, we make use of a three--state stochastic spin model (referred here as lattice Blume-Capel model, see Section~\ref{sto}) and an associated phase-field type model (a coupled system of nonlinear diffusion-drift equations analyzed in \cite{Lyons_NonAna_2024}). The results presented here are new and still in a preliminary phase; however, we plan to study the evaporation extensively elsewhere, possibly for the 3D case. 
Benefiting of our prior experience with alike settings (see, e.g., \cite{Andrea_EPJ,Andrea_PhysRevE,Mario,LyonsMunteanetal2023}), we  focus within this framework exclusively on  ``from-the-top evaporation", in which evaporation takes place with equal rates on the whole considered domain. Another very interesting case, the so-called ``from-the-side evaporation",
in which evaporation takes place only on one of the four sides of the 
domain, will be treated in detail elsewhere.  


    
As main modeling tool, we use three--state spin 
lattice  systems and study the dependence of the domain growth (like in \cite{cirillo2017sum}) on 
concentration, temperature, and initial composition. Interestingly, we are obtaining values of the 
growth exponents lower than the typical ones found for the binary models. 
This fact  suggests that the phase separation patterns that can form in such  
type of systems are complex. 

Our interest extends as well to exploring the 
ability of models based on partial differential equations (PDEs),  
obtained as continuum limit of three-state spin 
systems, to describe a similar formation of patterns. Our target PDE system is 
derived in the literature (see \cite{Marra}) as the rigorous hydrodynamic limit of a 
suitably scaled interacting particle system of Blume-Capel-type driven 
by Kawasaki dynamics. The system we have in mind describes the interaction within a 
ternary mixture that is the macroscopic counterpart of a mix of two 
populations of interacting solutes in the presence of a background solvent.  
We study the qualitative behavior of numerical simulations of  finite volume 
approximations of smooth solutions to our system and their quantitative 
post-processing in terms of the structure factor indicator.
We find  many qualitative features (e.g. general 
shape and approximate coarsening rates) similar to those observed earlier (cf. e.g. \cite{Andrea_PhysRevE}) on the 
stochastic Blume-Capel dynamics with three interacting species. 

The paper is organized in the following fashion. We introduce the reader to the topic 
of three-state stochastic spin models in Section~\ref{sto} and then illustrate numerically 
what morphologies can be usually expected as well as what typical effects at the 
morphology level are visible while varying the evaporation rate. The phase-field model we have 
in view here is introduced in Section~\ref{kacpde}. Similarly as in the previous section, 
we point out our numerical results on typical morphologies obtained now at the continuum 
level and show effects induced by changing the evaporation rate. 
We discuss briefly our findings in Section~\ref{discussion}.  

\section{The three--state stochastic spin model}
\label{sto}
Following the ideas firstly introduced in \cite{Andrea_EPJ}
and exploited in \cite{Andrea_PhysRevE}, we approach the phase separation question for interacting ternary mixtures 
using the Blume--Capel model, which we describe next for a two-dimensional scenario.
Consider $\mathbb{Z}^2$ embedded in $\mathbb{R}^2$ 
and refer to its elements as \emph{sites}.
Given two sites $i,i'\in\mathbb{Z}^2$, let $|i-i'|$ be their
Euclidean distance. 
Given $i\in\mathbb{Z}^2$, we say that 
$i'\in\mathbb{Z}^2$ is a \emph{nearest neighbor}
of $i$ if and only if $|i-i'|=1$.
Pairs of neighboring sites will be called \emph{bonds}.

Consider the torus $\Lambda=\{1,\dots,L\}^2\subset \mathbb{Z}^2$ and 
associate each site $i$ of $\Lambda$ with a spin variable $\sigma(i)$
taking values in the 
\emph{single spin state space} $\{-1,0,+1\}$.
We let 
$\mathcal{X}=\{-1,0,+1\}^{\Lambda}$
be the \emph{configuration} or \emph{state} space.

In the mathematics literature (see e.g., 
\cite{CO1996,cjs2024,cirillo2017sum}), 
the Hamiltonian of the Blume--Capel model is often written
as 
\begin{equation}
\label{eq:int000}
H(\sigma)
=
J\sum_{\langle i,j\rangle}[\sigma(i)-\sigma(j)]^2
-\lambda\sum_{i\in\Lambda}[\sigma(i)]^2
-h\sum_{i\in\Lambda}\sigma(i)
,
\end{equation}
for $\sigma\in\mathcal{X}$,
where the first sum is extended to the $2L^2$ bonds with periodic 
boundary conditions due to the torus topology.
The parameters $J>0$ and $\lambda,h\in\mathbb{R}$ are called 
\emph{coupling constant}, \emph{chemical potential}, 
and, respectively, \emph{magnetic field}. In the context of spin
models a site $i$ with 
$\sigma(i)=\pm1$ is said to be occupied by a particle 
with spin $\pm1$, while a site $i$ with $\sigma(i)=0$ is said to be empty.
In our polymer interpretation of the model the $\pm1$ and $0$ spins 
will represent, respectively, polymer and solvent molecules.

We remark that in the physics literature \cite{FGRN1994,blume1966} the 
Hamiltonian of the same model is usually written  as 
\begin{displaymath}
H(\sigma)=-\bar{J}\sum_{\langle i,j\rangle}\sigma(i)\sigma(j)
+D\sum_{i\in\Lambda}[\sigma(i)]^2
-h\sum_{i\in\Lambda}\sigma(i)
,        
\end{displaymath} 
with $D$ the \emph{crystal field}.
An easy computation yields that the two formulas are equivalent provided
$\bar{J}=2J$ and $D=4J-\lambda$.

\begin{figure}[ht]
\centering
\includegraphics[scale=.9]{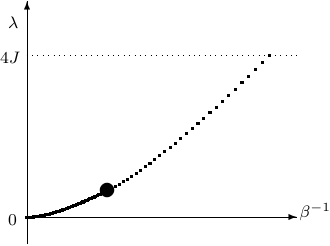}
\caption{Schematic representation of the phase diagram in the 
plane $\lambda$--$\beta^{-1}$,
namely, at $h=0$.
Thick solid and thick dotted lines represent, respectively, first and second 
order phase transition between the low temperature ordered and the high 
temperature paramagnetic phases. The small disk denotes the tricritical 
point whose coordinates, see, e.g., \cite{beale1986}, are estimated 
to be $(0.61,0.08J)$. The critical temperature at $\lambda=4J$ is 
$1.69$.}
\label{fig:phase_diagram}
\end{figure}

The equilibrium state of the model is provided by the usual 
Gibbs probability measure 
$\exp\{-\beta H(\sigma)\}/\sum_{\eta\in\mathcal{X}}\exp\{-\beta H(\eta)\}$,
where $\beta$ is the inverse of the temperature.
The Blume--Capel model phase diagram has been widely studied in the 
physically relevant region $D\ge0$, which means, with our parametrization,
$\lambda\le4J$. A very rich behavior has been found, see, e.g., \cite{beale1986} 
and references therein. Indeed, as shown in the schematic 
phase diagram reported in Fig.~\ref{fig:phase_diagram}, at 
$h=0$ the model exhibits a transition between an ordered and a disordered
paramagnetic phase that, depending on the value of $\lambda$, is first or
second order.

Within this context, we wish to exploit the ability of the model to 
show three-state pattern formation powering it with a stochastic 
dynamics which preserves the value of the spins.
A natural choice is the reversible, with respect to the 
Gibbs equilibrium measure, Kawasaki dynamics in which a 
bond is selected at random with uniform probability throughout the torus 
and the two spins associated with the two sites of the bond 
are swapped with probability one if the variation of energy 
$\Delta H$
due to the swap is non-positive or with probability 
$1-\exp\{-\beta \Delta H\}$ if $\Delta H>0$.

Indeed, starting from a completely random configuration, with 
prescribed content of zero, plus, and minus spin, the system, 
provided the temperature is small enough, will evolve forming
domains of constant spin values whose shape and size 
will strongly depend on the ratio of the spin mixture 
prescribed {\em a priori}. We denote, here and in the following, 
by $c_0$, $c_1$, and $c_{-1}=1-(c_0+c_1)$ the fixed 
fraction of zeros, pluses, 
and minuses, respectively.

This coarsening behavior is due to the structure of the Hamiltonian 
which favors configurations in which spins are  
surrounded by spins of the same values. The Blume--Capel model has the 
peculiarity that different interfaces have different energy costs, 
indeed, according to \eqref{eq:int000}, we have that the 
minus--plus interface cost $4J$, while the minus--zero and minus--plus 
interface cost is $J$. In view of describing morphologies 
of polymer--polymer--solvent mixtures, the interface costs suggest 
to interpret the zero spin as the solvent component and the 
minus and plus spins as the two polymer components.

As mentioned above, we are interested in the morphology formation 
under evaporation of the solvent component. Then, we have to modify 
the dynamics of the stochastic Blume--Capel model allowing the 
zero spins to abandon the lattice. 
This effect is introduced 
by drawing at random a site of the lattice 
at each step of the dynamics and, provided it is a zero, 
flipping it to minus with probability $\alpha c_{-1}/(c_{-1}+c_{+1})$ and 
to plus with probability $\alpha c_{+1}/(c_{-1}+c_{+1})$, 
where $0\le\alpha\le1$. Note that, 
the drawn zero spin has anyhow the probability $1-\alpha$ to be kept
at its place in the lattice. Thus, the parameter $\alpha$ controls the evaporation speed
in the sense that the lower is $\alpha$ the lower is the 
speed at which the evaporation of the zero component takes place.

\subsection{Choice of the parameters}
\label{sto-par}
As explained above and in Section~\ref{intro} we are interested in
using the stochastic Blume--Capel model to investigate the
different morphologies that can be formed due to phase separation.
These phenomena will obviously depend strongly on how the
parameters appearing in the Hamiltonian will be chosen.
But,
since we are using the Kawasaki dynamics, the number of minus, zero,
and plus spins is conserved during the time evolution, so that the
terms in $\lambda$ and $h$ in the Hamiltonian are constant and, hence,
the two related parameters are not relevant.
Thus in the simulations we will always set
$J=1$ and control the dynamics via the sole parameter $\beta$.

\begin{figure}[!ht]
\begin{picture}(450,170)(-2,0)
\put(5,0)
{
 \includegraphics[width=.45\textwidth,height=.45\textwidth]{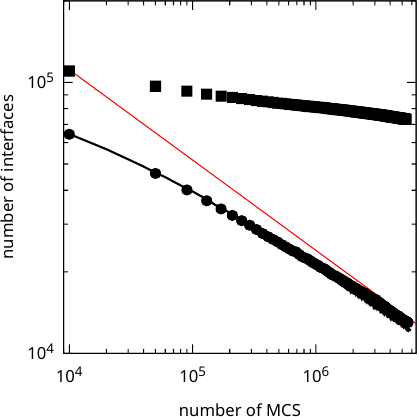}
}
\put(185,24)
{
 \includegraphics[width=.125\textwidth,height=.125\textwidth]{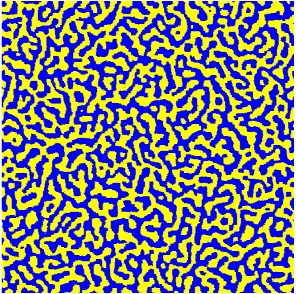}
}
\put(235,24)
{
 \includegraphics[width=.125\textwidth,height=.125\textwidth]{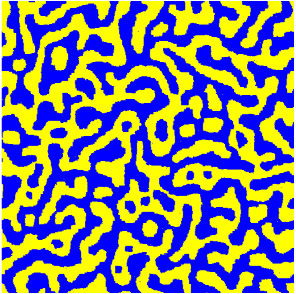}
}
\put(285,24)
{
 \includegraphics[width=.125\textwidth,height=.125\textwidth]{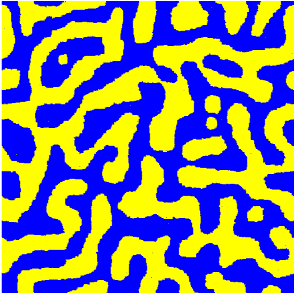}
}
\put(185,69)
{
 \includegraphics[width=.125\textwidth,height=.125\textwidth]{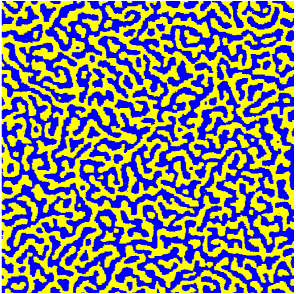}
}
\put(235,69)
{
 \includegraphics[width=.125\textwidth,height=.125\textwidth]{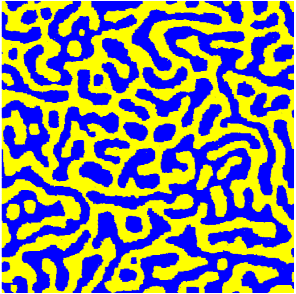}
}
\put(285,69)
{
 \includegraphics[width=.125\textwidth,height=.125\textwidth]{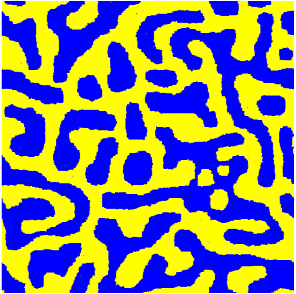}
}
\put(185,114)
{
 \includegraphics[width=.125\textwidth,height=.125\textwidth]{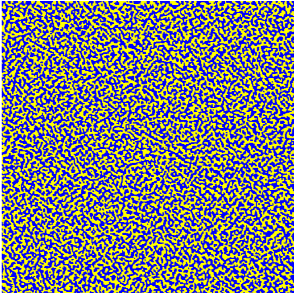}
}
\put(235,114)
{
 \includegraphics[width=.125\textwidth,height=.125\textwidth]{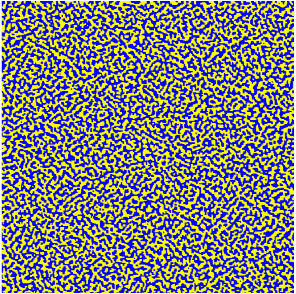}
}
\put(285,114)
{
 \includegraphics[width=.125\textwidth,height=.125\textwidth]{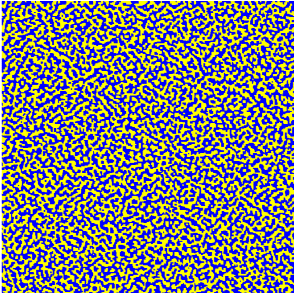}
}
\end{picture}
\caption{Number of plus-minus interfaces as function of the number of 
Monte Carlo Sweeps on the left and system configurations on the right
for a simulation on the lattice $512\times512$ with $J=1$ and
initial uniformly random configuration. For the Blume--Capel model
we used $c_0=0$.
\textbf{Left:}
The black line, the circles, and the squares
refer respectively to the Ising system with $\beta=1.1$ and
the Blume--Capel model with $\beta=0.55$ and $\beta=1.1$.
The red solid straight line is a reference line with slope $-1/3$.
\textbf{Right:} yellow and blue points represent, respectively, minus and
plus spins. From the bottom to the top: Ising $\beta=1.1$, Blume--Capel
$\beta=0.55$, and Blume--Capel $\beta=1.1$. From the 
left to the right MCS $=10^5, 10^6, 5\times10^6$.}
\label{fig:sto000}
\end{figure}

In the case in which the zero content $c_0$ is chosen equal to $0$,
the system looks like the ferromagnetic Ising model with coupling
constant $J$ with the slightly
difference that a plus--minus interfaces costs here $4J$ with
respect to the reference value $0$ for the homologous spin bonds,
whereas in the Ising model plus--minus
interfaces cost $J$ with respect to the reference value $-J$
for the homologous spin bonds. Thus, if we denote by $n'$ and
$n$ the number of interfaces after and before a spin swap
on a bond, we have that the change in energy $\Delta H_\textrm{bc}$
and $\Delta H_\textrm{i}$ respectively for the Blume--Capel and Ising model
are
\begin{displaymath}
\Delta H_\textrm{bc}=4J (n'-n)
\end{displaymath}
and
\begin{displaymath}
\Delta H_\textrm{i}=
J n'-(7-n')J
-[
  J n-(7-n)J
 ]
=2J(n'-n)
,
\end{displaymath}
where we have used that the number of bonds involved in a swap is
seven.

Since
$\Delta H_\textrm{bc}=2\Delta H_\textrm{i}$, we have that if the
Ising and the Blume--Capel dynamics are run with the same
$\beta$,  swaps against the energy drift in
the Blume--Capel model are highly less probable than in the
Ising system. This implies that the
Blume--Capel Kawasaki dynamics is highly slower than the
corresponding Ising one.
We have checked this fact running a simulation on the
lattice $512\times512$ for the Ising model with
$\beta=1.1$ and the Blume--Capel model with $\beta=1.1$ and $\beta=0.55$
with zero fraction $c_0=0$.
In Fig.~\ref{fig:sto000}, together with the configuration of the system, we have
plotted, as a function of the number of Monte Carlo Sweeps (MCS),
the number of plus--minus interfaces. It is well known that for the
Ising model, in the scaling regime, this number decreases
with a power low with exponent $1/3$. The number of interfaces 
has been obtained by measuring the value $E$ of the Hamiltonian of the 
time dependent configuration and by computing 
$E/4$ in the Blume--Capel simulation with $c_0=0$ and 
$(E+2L^2)/2$ in the Ising simulation. 

The picture shows quite neatly the scaling behavior of the
Ising model at $\beta=1.1$ and of the Blume--Capel model
at $\beta=0.55$, while the Blume--Capel model at higher $\beta$, after
a quick formation of minus and plus structures, appears completely frozen.

However, in the simulations that we will discuss in the following section, we shall always consider cases with $c_0 \neq 0$ and the presence of zero spin particles will highly speed up the dynamics since the energy cost of zero--plus and zero--minus interfaces is equal to $J$.
This phenomenon was already noted in \cite{Andrea_PhysRevE} and a similar effect was also observed in \cite{fratzlpenrose1994}, where a small density of lacunas was added to the Ising model and it was considered a modified Kawasaki dynamics in which only swaps between spins and lacunas were allowed. 

In the following, our simulations will be run
on the lattice $512\times512$ with parameters $J=1$ and $\beta=1.1$.

\subsection{Morphologies}
\label{sto-noev}
\begin{figure*}[h!]
    \centering
    \begin{subfigure}[t!]{0.4\textwidth}
        \centering        
        \includegraphics[height= \textwidth]{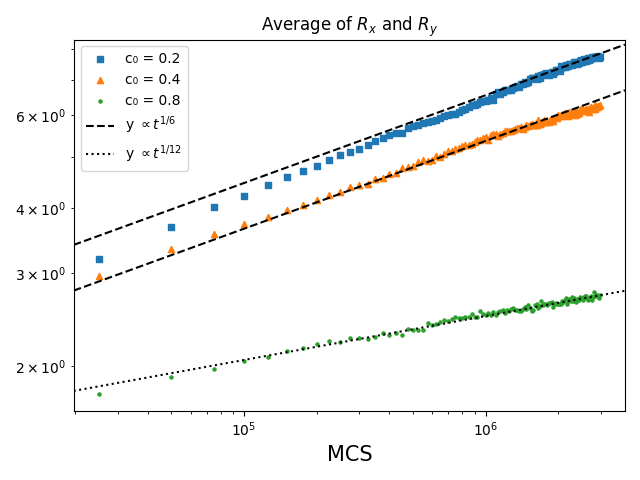}
\end{subfigure}%
    ~
    \begin{subfigure}[t!]{0.7\textwidth}
        \centering        
        \includegraphics[height=0.22\textwidth]{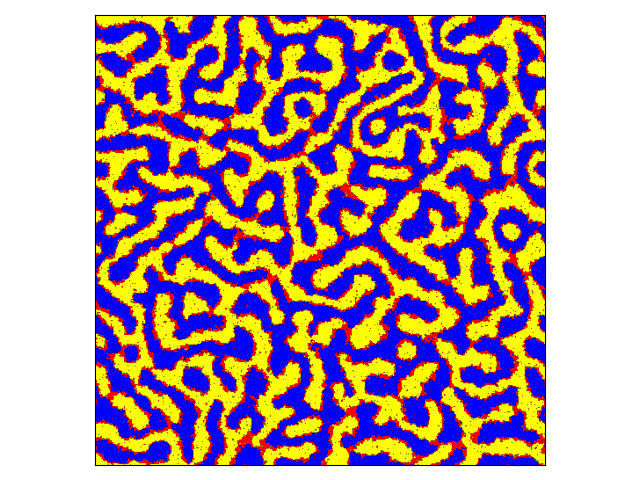}
        ~ %
        \includegraphics[height=0.22\textwidth]{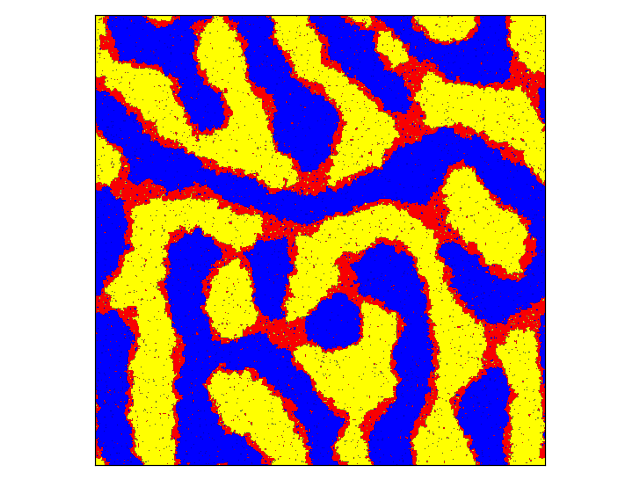}

        \includegraphics[height=0.22\textwidth]{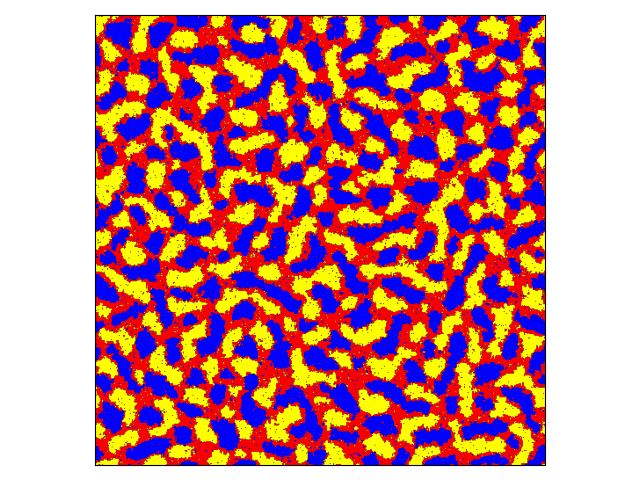}
        ~   
        \includegraphics[height=0.22\textwidth]{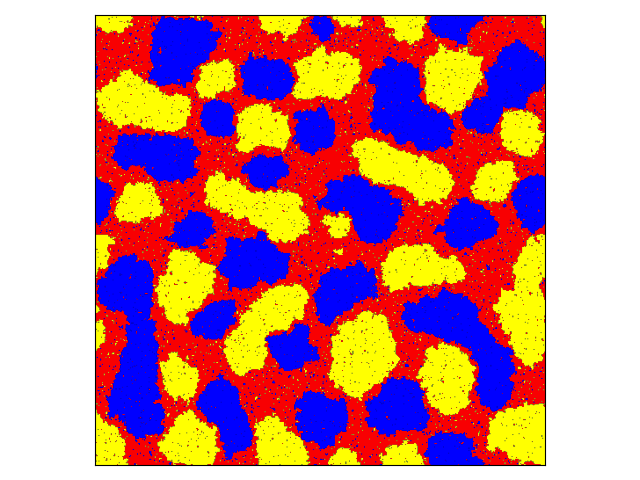}

        \includegraphics[height=0.22\textwidth]{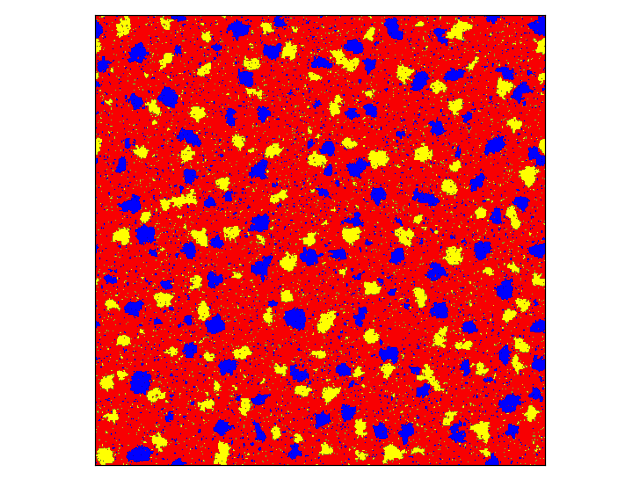}
        ~      
        \includegraphics[height=0.22\textwidth]{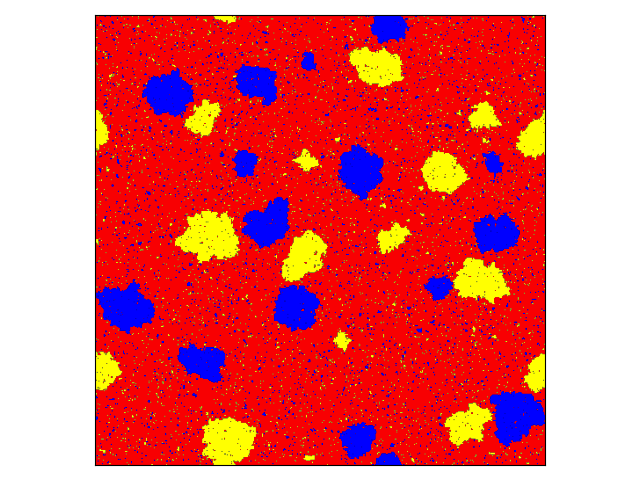}       
    \end{subfigure}%
    \caption{\textbf{Left:} Time evolution of the domain sizes given by the structure factor for the lattice model with reference lines $y = Ct^{p}$.  \textbf{Right:} Snapshots of the morphologies at early and final time steps (columns) for solvent ratio levels 0.2, 0.4, and 0.8 (rows). }
    \label{fig:noevap}
\end{figure*}

We discuss, here, the observed morphologies in the regime discussed 
above in presence of solvent, namely, 
when the zero content $c_0$ is not zero. In particular we shall 
consider the cases $c_0=0.2$, $0.4$, and $0.8$ reported in 
Fig.~\ref{fig:noevap}. We shall discuss qualitatively the 
configurations observed during the evolution, but we shall also provide
a quantitive characterization of the process 
by measuring the size of the growing domains 
by means of the inverse of the first moment of the structure factor. 
Namely, for any $(k_x,k_y) \in \{-\pi , -\pi + 2\pi/L, \dots, \pi - 2\pi/L, \pi\}^2$, we define the structure factor of a 2D configuration as
\begin{equation}\label{Eq:SF_Lattice}
    C((k_x,k_y),t) = \frac{1}{L^2} \left\vert \sum_{(x,y) \in \Lambda} \sigma_t(x,y)\, e^{i(k_x x + k_y y)}\right\vert^2.
\end{equation}
We can then estimate the horizontal and vertical size of the domains by defining 
\begin{equation}\label{Eq:SFR_Lattice}
    R_\alpha(t) = \frac{\sum_{(k_x,k_y)} C((k_x,k_y),t)}{\sum_{(k_x,k_y)} |k_\alpha | \, C((k_x,k_y),t)},
\end{equation}
where $\alpha \in \{x,y\}$.

For $c_0=0.2$, at all times, the minus and plus domains appear to be 
separated by a layer of zeroes and do not form a bicontinuous 
structure. It is remarkable that they  
have a strongly oblong shape during the whole evolution.
This results in a growth exponent smaller than the usual $1/3$ 
value typical of the two--state system and found in 
Section~\ref{sto-par} in the $c_0=0$ case. This is in agreement 
with what we found in \cite[bottom panel of Fig.~2]{Andrea_PhysRevE},
where we have show that the classical exponent $1/3$ is recovered 
in the case $c_0=0.1$ in a time interval in which the bicontinuos 
phase is observed in the lattice, see 
\cite[second row of Fig.~3]{Andrea_PhysRevE}.

In contrast, in the $c_0=0.8$ case the solvent background is so predominant 
that the plus and minus domains are small, almost spherical symmetrical balls,
growing slowly via a very complicated mechanism: minuses and pluses have to 
abandon their clusters (which is a highly improbable event), walk through the 
zero background, and be captured by a larger cluster. 
This mechanism is so slow that a smaller growth exponent is measured.

The case $c_0=0.4$ seems to be an intermediate situation, indeed, 
the domain are oblong at initial times, while tend to become 
spherically symmetric at larger times. Even the growth rate tends
to switch from the $c_0=0.2$ to the $c_0=0.8$ behavior.

\begin{figure*}[h!]
    \centering
    \begin{subfigure}[t!]{0.5\textwidth}
        \centering        
        \includegraphics[height= 0.8\textwidth]{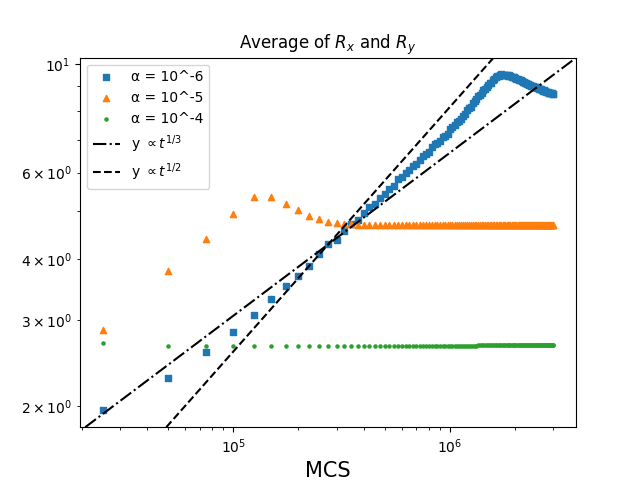}

\end{subfigure}%
    ~ 
    \begin{subfigure}[t!]{0.5\textwidth}
        \centering
        \includegraphics[height = 0.8\textwidth]{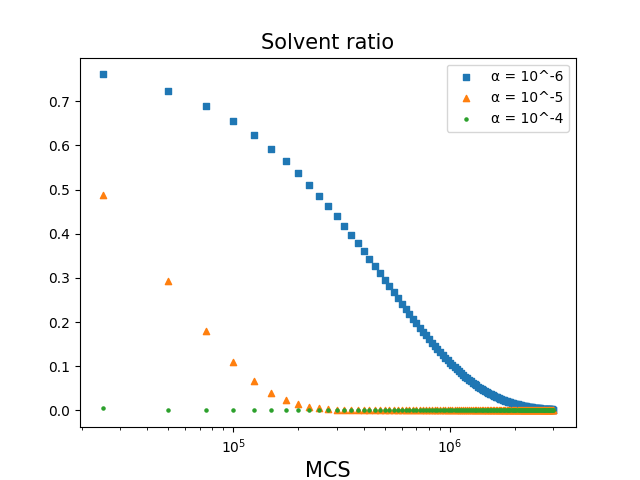}
    \end{subfigure}%
    
    \begin{subfigure}[t!]{\textwidth}   
        \centering        
        \includegraphics[height=0.22\textwidth]{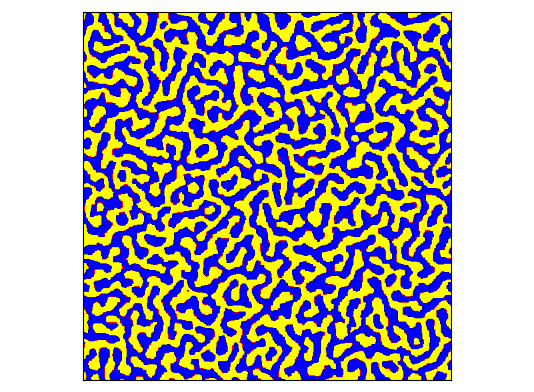}
        ~ 
        \includegraphics[height=0.22\textwidth]{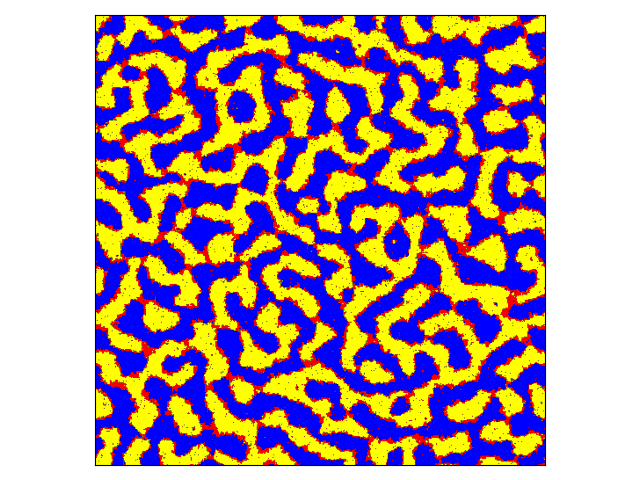}
        ~ 
        \includegraphics[height=0.22\textwidth]{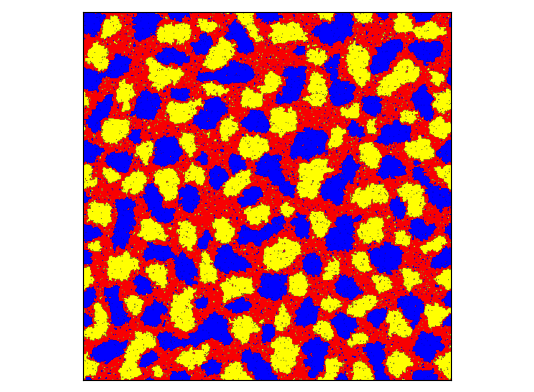}
             
        \includegraphics[height=0.22\textwidth]{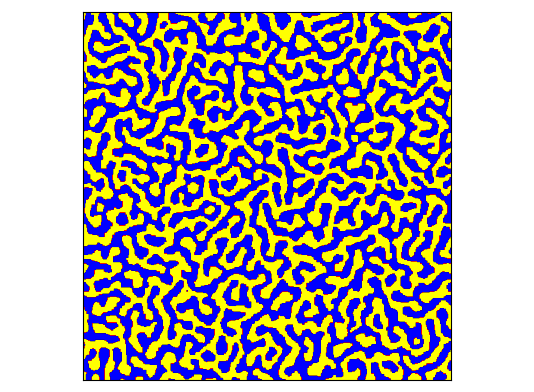}
        ~   
        \includegraphics[height=0.22\textwidth]{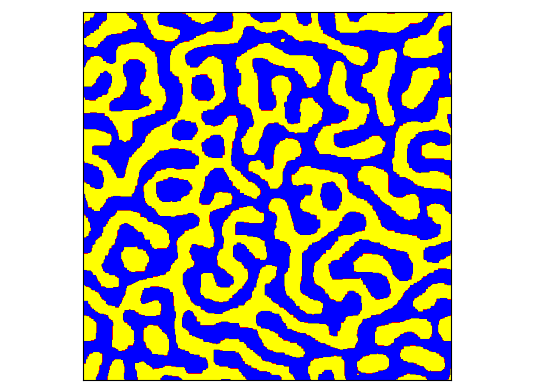}
        ~ 
        \includegraphics[height=0.22\textwidth]{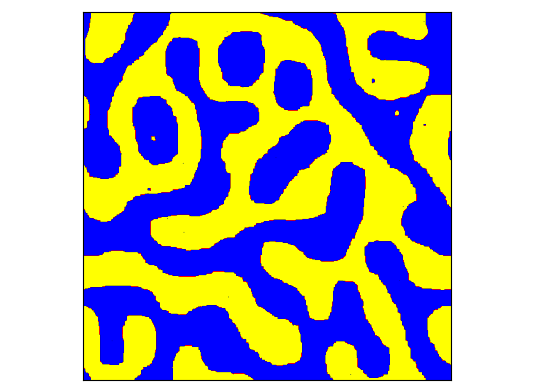}  
    \end{subfigure}%
    \caption{\textbf{Top left:} Time evolution of the domain sizes given by the structure factor for the lattice model with reference lines $y = Ct^{p}$.  \textbf{Top right:} Evolution of the solvent ratio throughout the simulation. \textbf{Bottom:} Snapshots of the morphologies at an early time (top row) and a late time (bottom row) for evaporation probabilities $\alpha = 10^{-4},\, 10^{-5},\, 10^{-6}$ (columns).}
    \label{fig:Lattice_evap}
\end{figure*}

\subsection{The effect of evaporation}
\label{sto-ev}
As explained in Section~\ref{sto} we consider a modified version 
of the stochastic model to implement the possibility of the 
evaporation of the solvent (zero) component. This is realized uniformly 
on the lattice and is controlled by the parameter $\alpha$ which 
tunes the evaporation rate.

In Fig.~\ref{fig:Lattice_evap} we report our findings for 
$\alpha=10^{-4}$, $10^{-5}$, and $10^{-6}$. In all the simulations 
discussed in this section the initial solvent content is $c_0=0.8$.

The case $\alpha=10^{-4}$ is absolutely trivial, indeed, evaporation is so fast 
that all the solvent is quicky removed from the lattice so that the 
system behaves as the stochastic Blume--Capel model with $c_0=0$. The 
dynamics, thus, as discussed in Section~\ref{sto-par}, freezes and no 
coarsening is observed in the time scale that we could simulate. 

The cases $\alpha=10^{-5}$ and $10^{-6}$ are, indeed, much more interesting. 
The initial time configurations are very similar to 
those that we observed in absence of evaporation at solvent 
concentration $c_0=0.4$. Nevertheless, the growth rate seems to be 
higher, with exponents close to the values $1/3$ and $1/2$ which 
are typical of the two--state systems, respectively, for conserved 
and non--conserved dynamics. This can be explained since in presence 
of evaporation, the dynamics is no more locally conserved, indeed 
spins are directly flipped from zero to plus or minus when an evaporation 
event happens. In some sense, 
the way in which we have introduced evaporation gives to the 
dynamics a sort of Glauber character.

In perfect agreement with what we observed in Section~\ref{sto-par} at 
later times, when the solvent component is fully evaporated, 
the dynamics freezes. The transition between the coarsening to the 
frozen regime happens, for $\alpha=10^{-5}$ and $10^{-6}$, with 
the appearance of a hump in the domain size plot. 
This can be related to the fact that the final layer of zeros 
separating minus and plus disappears and hence direct interfaces are produced. 

\section{The continuum model}
\label{kacpde}
The spin lattice model discussed in Section~\ref{sto} has several 
nice modelling features, such as its flexibility and the fact that
it is easy to be simulated. On the other hand, the way in which 
we have introduced the evaporation effect is rather brutal and not 
straightforward. It is then natural to try a different modelling 
strategy, switching from lattice to continuous space models.
This can be realized by considering a continuum limit of the Blume--Capel
model. This cannot be done with the nearest--neighbors interaction 
considered in Section~\ref{sto}, but a new version of the model with 
a Kac--type long range interaction must be considered. 

Following 
\cite{Marra} we let $\gamma$ be a positive real number and 
consider on $\mathcal{X}$ the Hamiltonian
\begin{equation}
\label{kac000}
H_\gamma (\sigma) 
= 
\frac{1}{2}\sum_{\substack{i \neq j \in \Lambda}}
J_\gamma (i-j)[\sigma(i)-\sigma(j)]^2 
-\lambda\sum_{i\in \Lambda} [\sigma(i)]^2 
-h\sum_{x\in \Lambda}\sigma(i), 
\end{equation}
where $J_\gamma: \mathbb{R}^2\to \mathbb{R}$
is a Kac potential function, i.e.,
\begin{equation}
\label{geigamma}
J_\gamma(r)=\gamma^d J(\gamma r)
\end{equation}
for all $r\in \mathbb{R}^2$, where 
$J\in C^2(\mathbb{R}^2)$ is such that $J(r)=J(-r)$ (symmetry), 
$\int_{\mathbb{R}^2} J(r)\mathrm{d}r=1$ (normalized to $1$), $J(r)=0$ 
if $|r|>1$ (supported in the unit ball). 
Note that $\gamma^{-1}$ is the range of the interaction.
We refer to the monograph 
\cite{Presutti} for more information on the context.

For the Kac version of the model it is possible to 
compute exactly the free energy in the so--called mean field 
limit $\gamma\to0$. This has been done in \cite{Marra} 
following the Lebowitz--Penrose approach introduced in \cite{lebowitzpenrose1966}.
Studying the mean field free energy it is possible to find 
a phase diagram which shares some similarities with the one reported in Fig.~\ref{fig:phase_diagram} for the original model.
For instance, the transition temperature at $\lambda=4$ is $1.701$ and
there is some evidence (see e.g., \cite{Marra}) of the existence of a tricritical point along the 
transition line close to $\lambda=0.55$ and $T=0.68$.

From our perspective, the main reason to consider the Kac version of the 
Blume--Capel model is that, in this framework, it is possible to derive
rigorously its continuum limit. Indeed, in \cite{Marra} the authors obtain the following continuum limit -- a system composed of two coupled non-local nonlinear parabolic equations 
\begin{eqnarray}\label{Eq:MainModel_withoutEvap}
    \partial_t m &=& \nabla \cdot \left[\nabla m - 2 \beta (\phi -m^2 ) (\nabla J * m) \right], \quad (t,x) \in (0,T)\times\Omega,\\
    \partial_t \phi &=& \nabla \cdot \left[ \nabla \phi - 2 \beta m (1 - \phi) (\nabla J * m) \right]+F(\phi).
\end{eqnarray}
Here, $t$ and $x$ represent the time and space variable. $m$ represents the 
average spin density (also called \emph{magnetization}), and $\phi$ represents the average 
squared spin density (also called \emph{solute volume concentration}). 
Additionally, $\Omega \, \subset \mathbb{R}^2$ is a cube with spatially periodic boundary conditions (i.e., homeomorphic to the unit torus) and $T$ is some positive time.

As a simple example, we assume that the evaporation rate is proportional to the solvent volume concentration, i.e., $F:\mathbb{R}\to \mathbb{R}$ defined by $F(r) = \alpha(1-r)$.
We prescribe the initial data
\begin{equation}
m(t=0)=m_0  \mbox{ and } \phi(t=0)=\phi_0 \mbox{ in } \bar\Omega.
\end{equation}
As noted in \cite{Marra}, if we set $u = (m,\phi)$, our system can be written for $F=0$ as a gradient flow structure given by 
\begin{equation}\label{Eq:GradStruct}
    \partial_t u = \nabla \cdot \left( M\nabla \frac{\delta \mathcal{F}}{\delta \vec{u}}\right).
\end{equation}
The mobility matrix is 
\[M = \beta (1-\phi) \begin{bmatrix}
    \phi + \frac{\phi^2 - m^2}{1-\phi} & m \\
    m & \phi
\end{bmatrix}\]
and the free energy functional $\mathcal{F}$ is given by
\[\mathcal{F}(u) = \int_\Omega f(\vec{u}) \diff{x} + \frac{1}{2} \int_\Omega \int_\Omega J(x-x') [m(x) - m(x')]^2 \diff{x'} \diff{x}, \]
where 
$f(u) = \phi - m^2
+ \beta^{-1} [\frac{1}{2}(\phi+m)\log(\phi+m) + \frac{1}{2}(\phi-m)\log(\phi-m)+ (1-\phi) \log(1-\phi) - \phi \log(2)]$. 
In general, if $F\neq 0$, then the gradient flow structure is usually lost. 
Interestingly, the structure could be regained if one would treat the from-the-side evaporation scenario. 

\subsection{Morphologies}
\label{pde-noev}
Mimicking the results presented in Section \ref{sto-noev}, we aim to measure the growth of the domains formed by the continuum model and qualitatively compare these dynamics to the lattice model.
We make use of the natural continuous analog of the structure factor \eqref{Eq:SF_Lattice} given by 
\begin{equation}\label{SF_Contin}
    C((k_x,k_y), t ) = \frac{1}{|\Omega|} \int_\Omega m(t,x,y) \, e^{i(k_x x + k_y y)} \diff{x},
\end{equation}
for $(k_x,k_y) \in [-\pi, \pi]^2$ and compute the horizontal and vertical size as 
\begin{equation}\label{SFR_Contin}
    R_\alpha (t) = \frac{\int_{[-\pi,\pi]^2} C(t,(k_x,k_y)) \diff{k_x} \diff{k_y}}{\int_{[-\pi,\pi]^2} |k_\alpha| C(t,(k_x,k_y))\diff{k_x} \diff{k_y}},
\end{equation}
where $\alpha \in \{x,y\}$.
We point out that in practice, we simulate the continuum model \eqref{Eq:MainModel_withoutEvap} with a finite volume scheme with a uniform mesh (similar to the one found in \cite{LyonsMunteanetal2023}) and these integrals are computed via Riemann sums which closely resemble the discrete counterparts \eqref{Eq:SF_Lattice} and \eqref{Eq:SFR_Lattice}. 

\begin{figure*}[!ht]
    \centering
    \begin{subfigure}[!ht]{0.4\textwidth}
        \centering        
        \includegraphics[height= \textwidth]{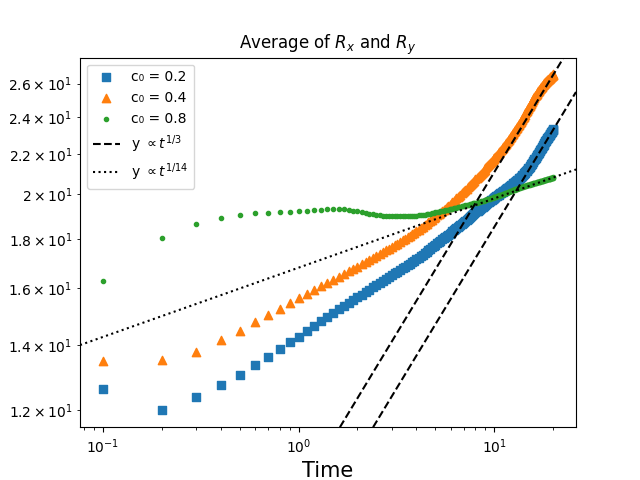}

\end{subfigure}%
    ~ 
    \begin{subfigure}[t!]{0.7\textwidth}
        \centering        
        \includegraphics[height=0.22\textwidth]{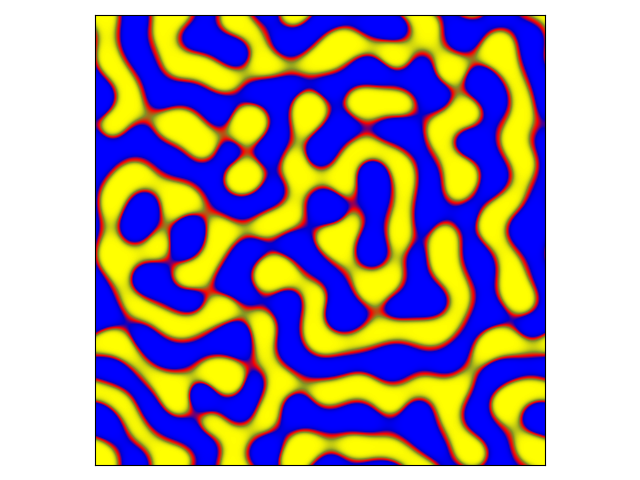}
        ~ %
        \includegraphics[height=0.22\textwidth]{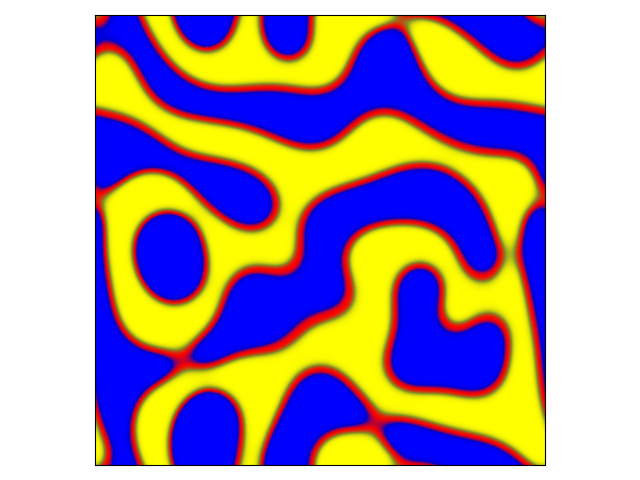}

        \includegraphics[height=0.22\textwidth]{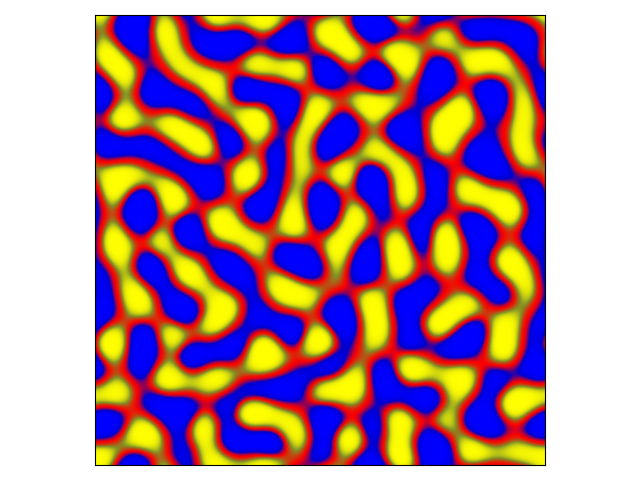}
        ~   
        \includegraphics[height=0.22\textwidth]{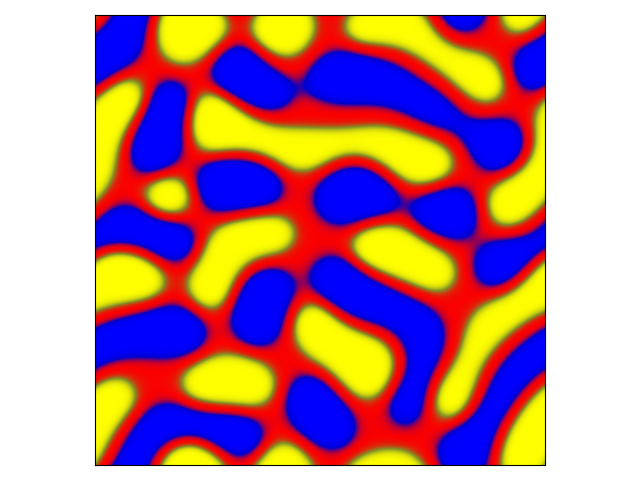}

        \includegraphics[height=0.22\textwidth]{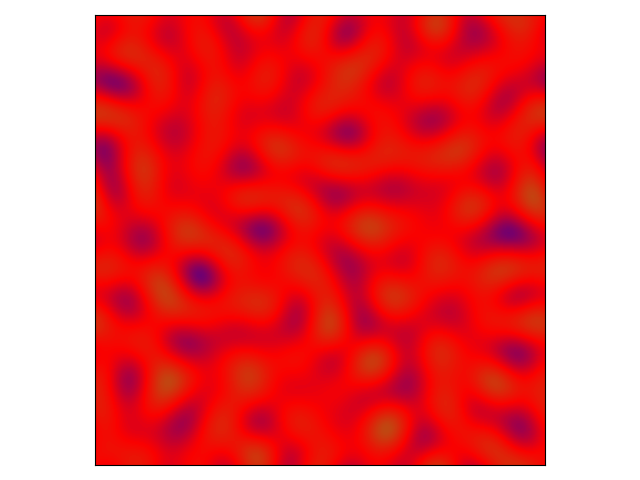}
        ~      
        \includegraphics[height=0.22\textwidth]{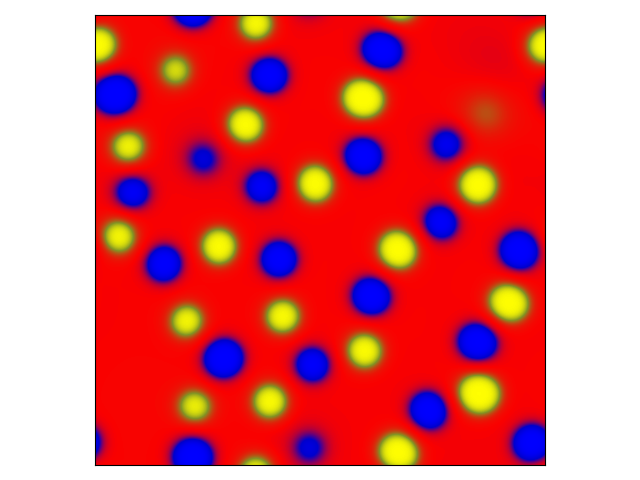}       
    \end{subfigure}%
    \caption{\textbf{Left:} Time evolution of the domain sizes given by the continuous analog of the structure factor.  
    \textbf{Right:} Snapshots of the morphologies at early ($t= 1$) and final ($t= 20$) time steps (columns) for solvent ratio levels 0.2, 0.4, and 0.8 (rows).}
    \label{fig:Continuum_noevap}
\end{figure*}

In Fig. \ref{fig:Continuum_noevap}, we plot the evolution of this continuous structure factor over time for the solvent levels $c_0 = 0.2, 0.4, $ and $0.8$. 
Additionally, we include snapshots of the configurations at an early time ($t= 1$) and the final time of the simulation.
Here and throughout the manuscript, we only plot the magnetization field, $m$, as the $\phi$ field is bounded from below by $|m|$ \cite[Theorem 1.1]{Lyons_NonAna_2024} and therefore is superfluous when the morphologies are well formed (see, however, \cite{LyonsMunteanetal2023} for configurations of the $\phi$ field). 

Generally, we observe similar qualitative behaviour to that seen in Fig. \ref{fig:noevap} for the lattice model. 
Indeed, in the $c_0 = 0.2$ case, long and strongly connected morphologies are generally observed with a layer of solvent particles separating the +1 and -1 phases. 
We likewise see similarities with the lattice model in the structure of  $c_0 = 0.4$ and $c_0 = 0.8$ cases.

Additionally, we see that there are some similarities with regards to the coarsening rates as well. 
For instance, the case $c_0 = 0.8$ displays the slowest growth rate due to the phases being isolated in spherical clusters. 
The $c_0 = 0.4$ case again acts as a transition regime where the coarsening rate slows down towards the end of the simulation where the phases become more isolated. 
The difference in the two models appears to be more quantitative as a 1/3 power law is achieved during certain time regimes.

\subsection{The effect of evaporation}
\label{pde-ev}

Making use of the linear evaporation relation, $F(\phi) = \alpha (1-\phi)$, we can model evaporation in a similar way to that done in Section \ref{sto-ev}.

\begin{figure*}[!ht]
    \centering
    \begin{subfigure}[t!]{0.5\textwidth}
        \centering        
        \includegraphics[height= 0.8\textwidth]{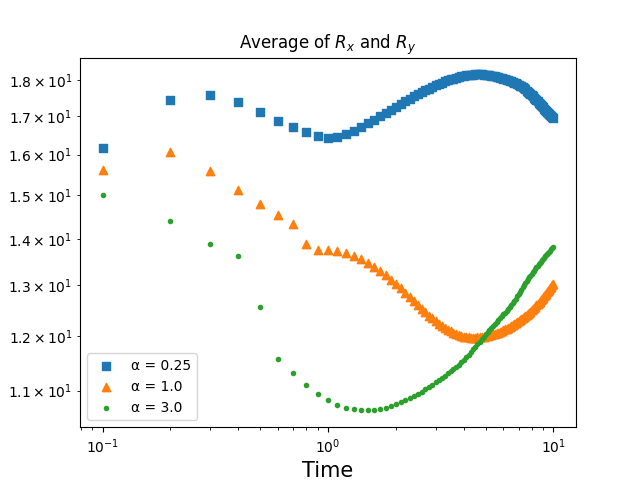}

\end{subfigure}%
    ~ 
    \begin{subfigure}[t!]{0.5\textwidth}
        \centering
        \includegraphics[height = 0.8\textwidth]{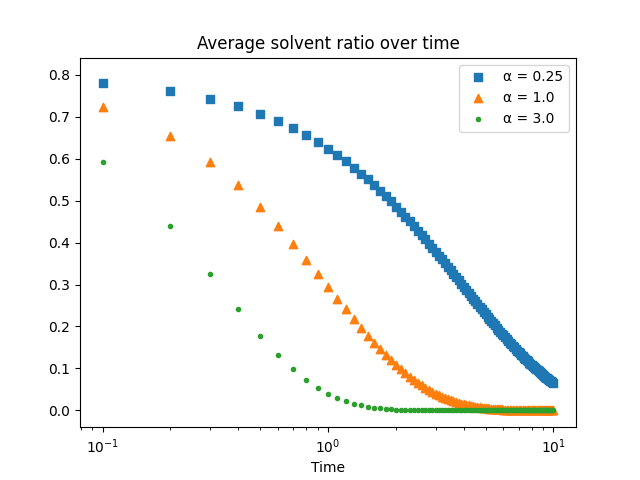}
    \end{subfigure}%
    
    \begin{subfigure}[t!]{\textwidth}   
        \centering        
        \includegraphics[height=0.22\textwidth]{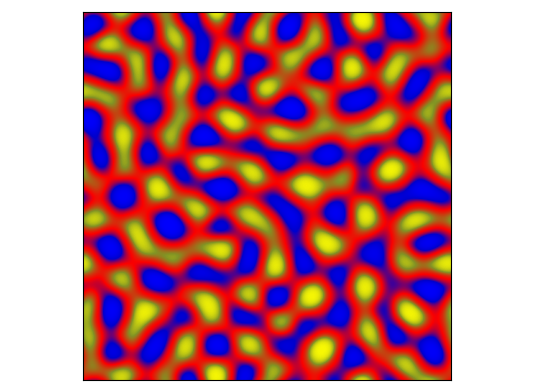}
        ~ 
        \includegraphics[height=0.22\textwidth]{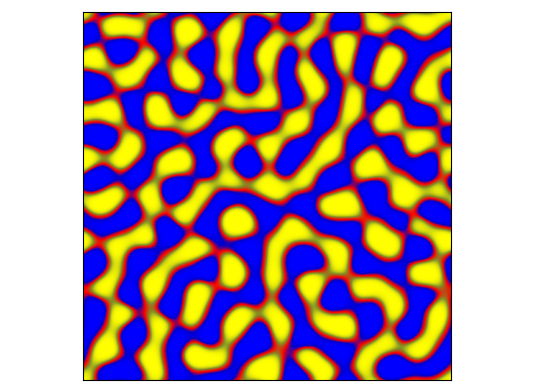}
        ~ 
        \includegraphics[height=0.22\textwidth]{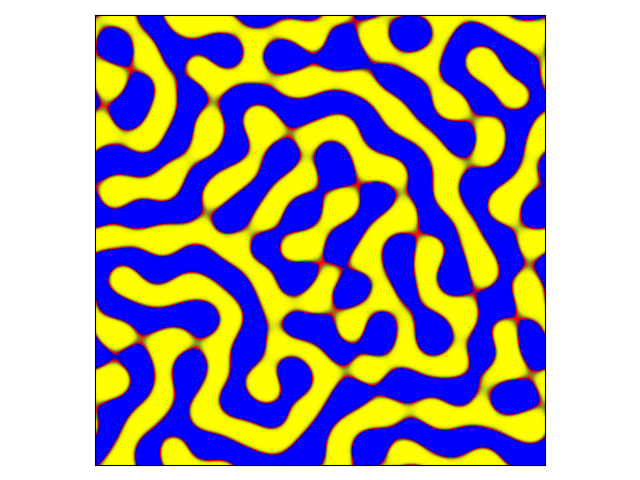}
             
        \includegraphics[height=0.22\textwidth]{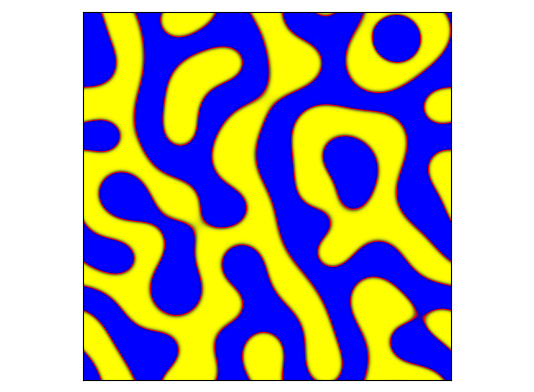}
        ~   
        \includegraphics[height=0.22\textwidth]{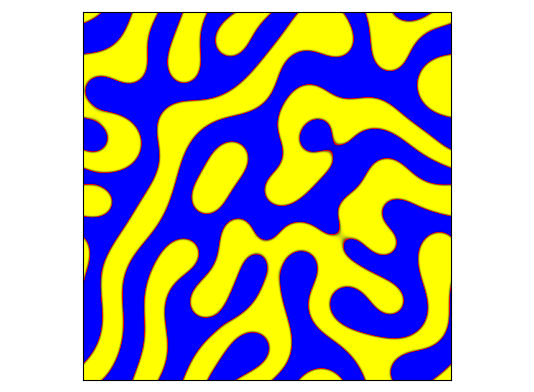}
        ~ 
        \includegraphics[height=0.22\textwidth]{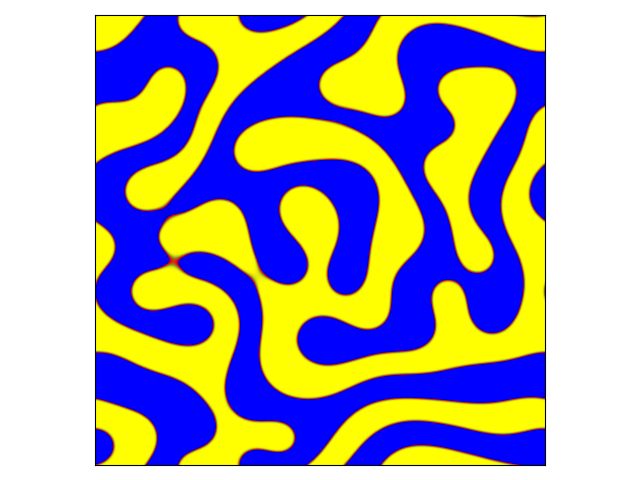}  
    \end{subfigure}%
    \caption{\textbf{Top left:} Time evolution of the domain sizes given by the structure factor for the continuum model.  \textbf{Top right:} Evolution of the solvent ratio throughout the simulation. \textbf{Bottom:} Snapshots of the morphologies at an early time (top row), $t = 1$, and the final time (bottom row), $t = 10$, for $F(\phi) = \alpha(1-\phi)$ and $\alpha = 0.25, \, 1, \, 3$ (columns).}
    \label{fig:Continuum_evap}
\end{figure*}

\begin{figure*}[!h]
    \centering
    \includegraphics[scale = 0.2]{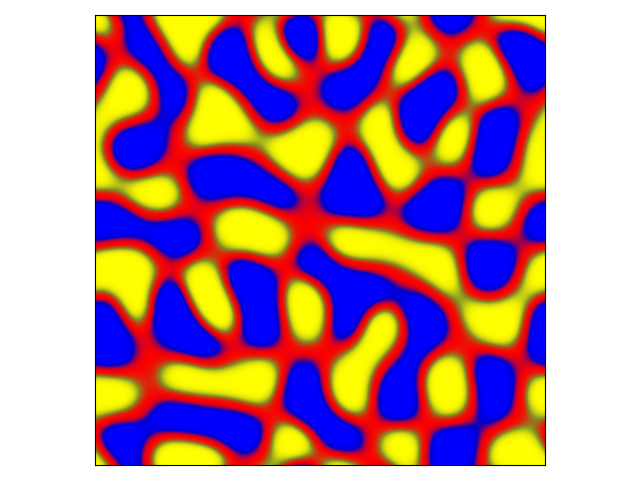}
    \includegraphics[scale = 0.2]{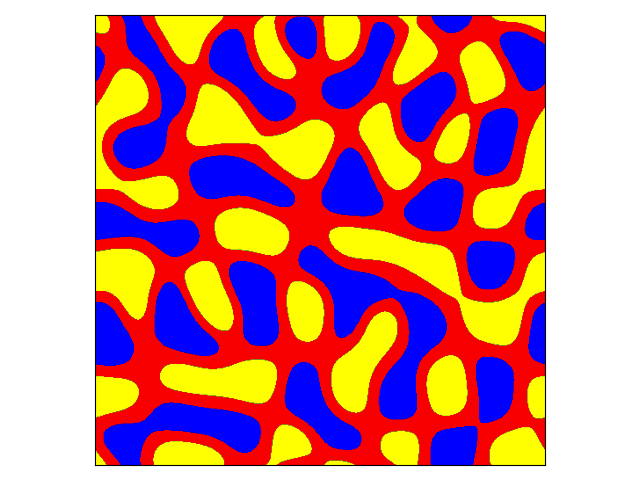}
        \caption{\textbf{Left:} Example of a possible configuration generated by the continuum model \eqref{Eq:MainModel_withoutEvap}.  
        \textbf{Right:} Rounded representation of the configuration.}
    \label{fig:Rounding_Example}
\end{figure*}

Since the use of indicators (like the structure factor) in continuum models appears to be nonstandard in the literature, we opt to explore measuring the domain sizes in multiple ways.
In Fig.~\ref{fig:Continuum_noevap} and \ref{fig:Continuum_evap}, we use the natural continuous analog of the structure factor indicator, however, we see that in the case of evaporation, e.g., Fig. \ref{fig:Continuum_evap}, the results are non-intuitive.
We explore measuring the domain sizes by first representing the continuum morphologies with a discrete configuration by rounding the phases to the nearest integer, essentially creating a type of sharp interface limit. 
In Fig. \ref{fig:Rounding_Example}, we display an example of this representation.

\begin{figure*}
    \centering
    \begin{subfigure}[t!]{0.5\textwidth}
        \centering        
        \includegraphics[height= 0.8\textwidth]{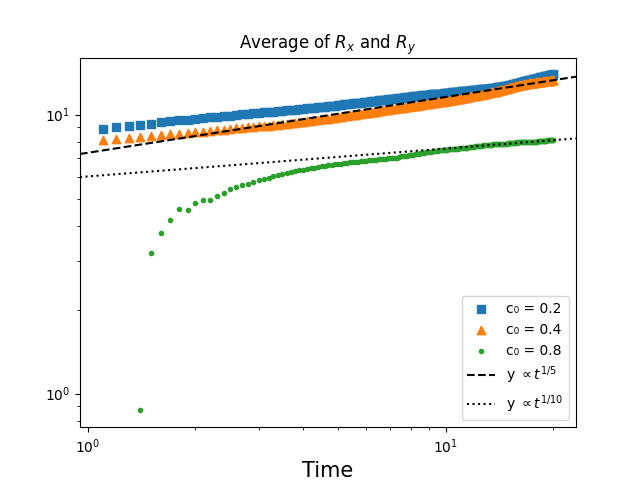}

    \end{subfigure}%
    ~ 
    \begin{subfigure}[t!]{0.5\textwidth}
        \centering
        \includegraphics[height = 0.8\textwidth]{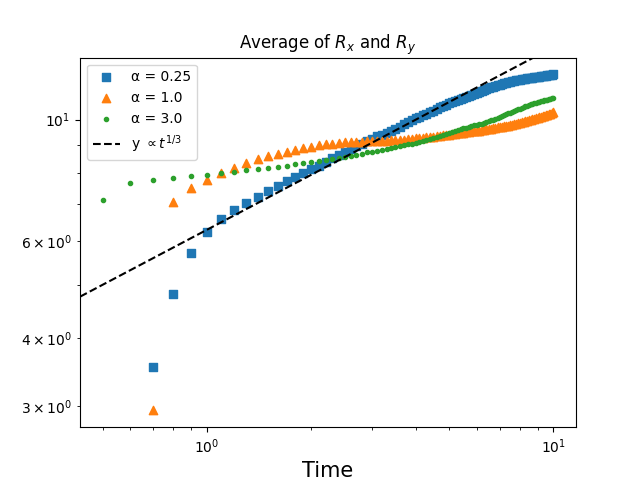}
    \end{subfigure}%
        \caption{\textbf{Left:} Structure factor calculations for the sharp interface representation of the configurations used in Fig. \ref{fig:Continuum_noevap}.  
        \textbf{Right:} Structure factor calculations for the sharp interface representation of the configurations used in Fig. \ref{fig:Continuum_evap}.}
    \label{fig:Rounded_SF}
\end{figure*}

We can then use the discrete structure factor indicator from Section \ref{sto-noev}. 
In Fig. \ref{fig:Rounded_SF}, we plot the evolution of the structure factor indicator in the absence and presence of evaporation as a function of time on these rounded representations. 
In both cases, the domains are generally smaller as the rounding process attributes more of the phases to the solvent phase. 
However, there is a significant difference in the coarsening rates of the configurations especially in the case of evaporation.
Here, instead of the oscillating behavior seen in Fig. \ref{fig:Continuum_evap}, we see a much more intuitive size trajectory. 

\section{Conclusion}\label{discussion}
We presented succinctly two conceptually different models that can capture phase separation in ternary mixtures with one evaporating component from two different perspectives -- {\em microscopic} (the three-state stochastic spin model acting on a lattice) and {\em macroscopic} (the continuum model in terms of partial differential equations). The lattice model is versatile in tuning the interaction parameters and is easier 
    to simulate numerically. On the other hand, modeling evaporation in the 
    lattice model is more tricky, while at the continuum level this presents no difficulties. Having direct access to a large class of morphologies produced computationally with either of the models, can serve as a starting platform for further investigations, such as a detailed study of the transport of charges through such heterogeneous materials or the study of mechanical properties of such materials. 

\bigskip
\par\noindent
\textbf{Acknowledgments}
R.L. and A.M. are grateful to Carl Tryggers Stiftelse for their financial support via the grant CTS 21:1656. S.A.M. would like to acknowledge the funding from the Swedish National Space Agency (Rymdstyrelsen), grant 2021-00137. ENMC thanks the PRIN 2022 project
``Mathematical modelling of heterogeneous systems"
(code 2022MKB7MM, CUP B53D23009360006). 
SAM and AM thank the National Academic Infrastructure for Supercomputing
in Sweden (NAISS) at the National Supercomputer Centre (NSC) at Linköping
University, partially funded by the Swedish Research Council through Grant
Agreement Nos. 2022-06725 and 2018-05973.

\end{document}